\documentstyle[11pt]{article} 
\newcommand{\beq}[1]{\begin{equation}\label{#1}}
\newcommand{\eeq}{\end{equation}}
\newcommand{\bear}[1]{\begin{eqnarray}\label{#1}}
\newcommand{\ear}{\end{eqnarray}}
\newcommand{\nn}{\nonumber}

\newcommand{\e}{\mbox{\rm e}}
\newcommand{\R}{\mbox{\rm I$\!$R} }
\newcommand{\p}{\partial}
\newcommand{\eps}{\varepsilon }
\newcommand{\btd}{\bigtriangledown}
\newcommand{\btu}{\bigtriangleup}
\newcommand{\rank}{\mbox{\rm rank} }
\newcommand{\diag}{\mbox{\rm diag} }
\newcommand{\abs}[1]{\vert{#1}\vert}
\newcommand{\foom}[1]{\protect\footnotemark[#1]}
\newcommand{\email}[2]{\footnotetext[#1]{e-mail: #2}}%
\newcommand{\Title}[1]{\noindent {\Large #1} \\}

\begin{document}
\begin{center}
\Title{\large\bf INTEGRABLE SPHERICALLY SYMMETRIC P-BRANE MODELS\\
ASSOCIATED WITH LIE ALGEBRAS}

\bigskip

\noindent{\normalsize\bf
S. Cotsakis\foom 1\dag
V. R. Gavrilov\foom 2\ddag,
V. N. Melnikov\foom 3\ddag}

\medskip

\noindent{\dag\ \it
Research Laboratory for Geometry, Dynamical
Systems and Cosmology (GEODYSYC), Department of Mathematics
University of the Aegean, Karlovassi, 83 200 Samos, Greece}

\medskip

\noindent{\ddag\ \it Centre for Gravitation and Fundamental Metrology
\\
VNIIMS, 3-1 M. Ulyanovoy St., Moscow 117313, Russia}

\vspace{5mm}
{\bf Abstract}
\end{center}
A classical model of gravity theory with several  dilatonic
scalar fields and differential forms admitting an interpretation in terms
of intersecting $p$-branes is studied
in (pseudo)-Riemannian space-time manifold
${\rm M} = \R_+  \times S^{d_0} \times \R_t\times M_2^{d_2}
\ldots \times M_n^{d_n},\ \ (n\geq 2)$  of dimension $D$.
The equations of motion of the
model are reduced to the Euler-Lagrange equations for the so-called
pseudo-Euclidean Toda-like system. We suppose that the
characteristic vectors related to the configuration of $p$-branes
and their couplings to the dilatonic scalar fields may be interpreted as
the root vectors of a Lie algebra of the types $A_r,\ B_r,\ C_r$.
In this case the
model is reducible  to  one of the open Toda chain's algebraic
generalization and is completely integrable by the known methods.
The corresponding general solutions are presented in explicit form.  The
particular exact solution describing a class of nonextremal black holes is
obtained and analyzed.

\email 1 {skot@aegean.gr}
\email 2 {gavr@rgs.phys.msu.su}
\email 3 {melnikov@rgs.phys.msu.su}

\section{Introduction}
There has been much interest in black hole solutions in $p$-brane theory
\cite{Stelle}-\cite{I9} because of the possible resolution of various
puzzles associated with quantum gravity\cite{Maldacena}.
A growth of interest in classical
$p$-brane solutions of  supergravities of various dimensions
is inspired by a conjecture that $D=11$ supergravity is a
low-energy effective field theory of eleven-dimensional fundamental
$M$-theory, which (together with so called $F$-theory) is a candidate for unification
of five known ten-dimensional superstring models.
Classical $p$-brane solutions may be considered as an instrument for
investigation of interlinks between superstrings and $M$-theory.

In this paper we consider generalized bosonic sector (without Chern-Simons
terms) of supergravity theories \cite{Cremmer} in the form of
a multidimensional gravitational model with several dilatonic scalar
fields and differential forms of various ranks admitting an
interpretation in  terms of intersecting $p$-branes.
As was shown in \cite{I3}, for cosmological and static
spherically symmetric space-times the equations of motion of such a model
are reduced to the Euler-Lagrange equations for the so-called
pseudo-Euclidean Toda-like Lagrange system. We reproduce this result in
Sec. 2.
Methods for integrating  of pseudo-Euclidean Toda-like systems
(see, \cite{G} and references therein)
are based on a Minkowski-like geometry for the
characteristic vectors determining the potential of the pseudo-Euclidean
Toda-like system. If the characteristic vectors form an orthogonal set,
then the pseudo-Euclidean Toda-like system is integrable. The corresponding
$p$-brane models have been studied in
\cite{I3},\cite{BKR},\cite{Br},\cite{I7}.
Here we apply these methods for integrating
the  $p$-brane model reducible to the algebraic generalizations of
an open Toda chain.  The characteristic vectors of such models may be
interpreted as  root vectors of the semi-simple Lie algebra.
We consider the Lie algebras of the  Cartan types $A_r,\ B_r,\ C_r$.
Using the technique suggested by Anderson \cite{Anderson} for
solving the Toda chain's equations of motion, in Sec. 3,4 we integrate
the $p$-brane models.
In the last section of the paper we examine the metric obtained for
some particular exact solution appearing for a quite wide class of
$p$-brane models.
This solution describes the nonextremal black hole under
some condition. The corresponding ADM-mass and the Hawking temperature
of a black hole are calculated.

\section{\bf The general model} \setcounter{equation}{0}

Following the papers \cite{I3},\cite{BIM},\cite{I7}
we consider here a classical
model of gravity theory with several dilatonic scalar fields
$\varphi^{\alpha}$ and differential
$n_a$-forms $F^a_{M_1\ldots M_{n_a}}$ in (pseudo)-Riemannian
space-time manifold  $ \rm M$ of  dimension $D$.
The action of the model reads
\beq{2.1}
S = \int\limits_{\rm M} d^{D}z \sqrt{|g|}
\left(
{R}[g] -
\sum_{\alpha,\beta=1}^\omega h_{\alpha\beta}\; g^{MN} \partial_{M}
\varphi^\alpha \partial_{N} \varphi^\beta 
- \sum_{a \in \Delta}
\frac{\e^{2 \lambda_{a} (\varphi)} }{n_a!}(F^a)^2 \right),
\eeq
where $ds^2 = g_{MN} dz^{M}dz^{N}$ is the
metric with Lorentzian signature on the manifold ${\rm M}$ ($M,N =0,1
\ldots, D-1$), $|g| = |\det (g_{MN})|$. $(h_{\alpha\beta})$ is a
symmetrical positively definite
$\omega\times \omega$ matrix, $\lambda_{a}
(\varphi)$ is a linear combination of the scalar fields, i.e.
\beq{2.2}
\lambda_a (\varphi)=\sum_{\alpha=1}^\omega
\lambda_{a,\alpha}\varphi^{\alpha},
\eeq
where $\lambda_{a,\alpha}$ are the coupling constants.
Furthermore
\bear{2.3} F^a = \frac{1}{n_a!} F^a_{M_1 \ldots M_{n_a}} dz^{M_1}
\wedge \ldots \wedge dz^{M_{n_a}}=dA^a ,\\ \label{2.4} A^a=
\frac{1}{(n_a-1)!} A^a_{M_1 \ldots M_{n_a-1}}
dz^{M_1} \wedge \ldots \wedge dz^{M_{n_a-1}} ,\\
\label{2.5}
(F^a)^2 =
F^a_{M_1 \ldots M_{n_a}} F^a_{N_1 \ldots N_{n_a}}
g^{M_1 N_1} \ldots g^{M_{n_a} N_{n_a}}.
\ear
The field  $A^a_{M_1,\ldots,M_{n_a-1}}$
may be called a gauge potential corresponding to the field strength
$F^a_{M_1,\ldots,M_{n_a}}$.  By $\Delta$  we denote some finite
set.

The action (\ref{2.1}) leads to the following equations of motion
\bear{2.6}
&& R_{MN} - \frac{1}{2} g_{MN} R  =   T_{MN},
\\
\label{2.7}
&& {\btu}[g] \varphi^\alpha =
\sum_{a \in \Delta}\frac{1}{n_a!}\lambda_a^{\alpha}
\e^{2 \lambda_{a}(\varphi)} (F^a)^2,\ \ \alpha=1,\ldots,\omega,
\\
\label{2.8}
&& \btd_{M_1}[g] (\e^{2 \lambda_{a}(\varphi)}
F^{a, M_1 \ldots M_{n_a}})  =  0,\ \ a\in \Delta.
\ear
The right side of the Einstein equations (\ref{2.6}) looks as follows
\bear{2.9}
T_{MN} =   T_{MN}[\varphi]+T_{MN}[F],
\ear
where we denoted
\bear{2.10}
&&T_{MN}[\varphi] = \sum_{\alpha,\beta=1}^\omega h_{\alpha\beta}
\left(\p_{M} \varphi^\alpha \p_{N} \varphi^\beta -
\frac{1}{2} g_{MN} \p_{P} \varphi^\alpha \p^{P} \varphi^\beta\right),\\
\label{2.11}
&&T_{MN}[F] =
\sum_{a\in\Delta}\frac{\e^{2 \lambda_{a}(\varphi)}}{n_{a}!}
\left( - \frac{1}{2} g_{MN} (F^a)^{2}
+ n_{a}  F^{a}_{M M_2 \ldots M_{n_a}} F_{N}^{a, M_2 \ldots M_{n_a}}\right).
\ear
In (\ref{2.7}),(\ref{2.8})
we denoted the Laplace-Beltrami operator and covariant derivative
with respect to the metric $g_{MN}$ by ${\btu}[g]$ and ${\btd}_M[g]$,
respectively.
The constants
$\lambda^{\alpha}_a$ in (\ref{2.7}) are introduced by
\beq{2.12}
\lambda^{\alpha}_{a} = \sum_{\alpha,\beta=1}^\omega h^{\alpha \beta}
\lambda_{a,\beta},
\eeq
where $(h^{\alpha \beta})$ is the inverse matrix to
$(h_{\alpha \beta})$.

Consider the model introduced under the following assumptions. Let the
\linebreak
$D$-dimensional space-time {\rm M} be decomposed into the direct product
of $\R_+$ (corresponding to a radial coordinate $u$), $d_0$-dimensional
sphere $S^{d_0}$ ($d_0\geq 2$), time axis $\R_t$ and $(n-1)$ factor spaces
$M_2, \ldots, M_n$, i.e.
\beq{2.13} {\rm M} = \R_+  \times S^{d_0} \times \R_t\times M_2^{d_2}
\ldots \times M_n^{d_n},\ \ n\geq 2.
\eeq
The metric on M is assumed correspondingly to be
\beq{2.14}
ds^2 = \e^{2{\gamma}(u)} du^2 + \e^{2x^0(u)}d\Omega_{d_0}^2 -
\e^{2x^1(u)}dt^2 + \sum_{i=2}^{n} \e^{2x^i(u)} ds_i^2,
\eeq
where $u$ is the radial coordinate,
$d\Omega_{d_0}^2=g^0_{m_0n_0}(y_0)dy_0^{m_0}dy_0^{n_0}$ is the line
element on $d_0$-dimensional unit sphere, $t$ is the time coordinate,
$ds_i^2=g^i_{m_in_i}(y_i)dy_i^{m_i}dy_i^{n_i}$ is the positively
definite metric on the $d_i$-dimensional factor space $M_i$,
$\gamma(u),x^2(u),\ldots,x^n(u)$ are scalar functions of the radial
coordinate $u$.
Herein, for reasons of simplicity, only Ricci-flat spaces
$M_2,\ldots,M_n$ are assumed (i.e. the components of the Ricci tensor for
the metrics $g^i_{m_in_i}$ are zero).

It is useful to consider $S^{d_0}$ and $\R_t$ as factor spaces $M_0$ and
$M_1$, respectively. So we put
\bear{2.15}
M_0\equiv S^{d_0},\\
M_1\equiv\R_t,\ d_1=1.
\ear
We split the coordinates on M
into the following ranges:
\beq{2.17}
\left(
z^0,z^1,\ldots,z^{d_0},z^{d_0+1},\ldots,z^{D-d_n},\ldots,z^{D-1}
\right)
=
\left(
u,y^1_{0},\ldots,y^{d_0}_{0},t,\ldots,y^{1}_{n},\ldots,y^{d_n}_{n}
\right).
\eeq
We introduce the following $d_i$-forms on M
\beq{2.18}
\tau_1=dt,\ \tau_i=\sqrt{\det(g^i_{m_in_i})}
dy_i^1 \wedge \ldots \wedge dy_i^{d_i},\ \ i=0,2,\ldots,n.
\eeq
Clearly, the canonical projection $\hat p_i:{\rm M}\to M_i$ of
$\tau_i$ provides with the volume form of $M_i$.

In order to construct the $p$-brane worldvolumes we introduce
submanifolds of the following type
\beq{2.19}
M_I=M_{i_1}\times\ldots\times M_{i_r},
\eeq
where
\beq{2.20}
I = \{ i_1, \ldots, i_r \},\ \ i_1 < \ldots < i_r,
\eeq
is any ordered non-empty subset of natural numbers $2,\ldots,n$.
Let $\Omega_0$ be the set of all such elements including the empty set,
i.e.
\beq{2.21}
\Omega_0= \{\emptyset,\{ 2 \}, \{ 3 \}, \ldots, \{ n \}, \{ 2, 3 \},
\ldots, \{ 2, 3, \ldots, n \} \}.
\eeq
By definition, put
\beq{2.22}
\bar I\equiv \{ 2, \ldots, n \}   \setminus I.
\eeq

In this paper we consider electrically charged $p$-branes with the
following worldvolumes
\beq{2.23}
M_I^{(e)}=\R_t\times M_I,\  I=\{i_1,\ldots,i_r\}\in \Omega_0.
\eeq
For empty $I=\emptyset$ we put $M_I^{(e)}=\R_t$.
The dimension of $M_I^{(e)}$ is given by
\beq{2.24}
d(I)\equiv\dim{M_I^{(e)}}=1+d_{i_1}+\ldots+d_{i_r}.
\eeq
($d(I)=1$ for $I=\emptyset$). The canonical projection
$\hat p_I:{\rm M}\to M_I^{(e)}$ of the following $d(I)$-form
\beq{2.25}
\tau(I)= dt\wedge\tau_{i_1} \wedge \ldots \wedge \tau_{i_r}
\eeq
is the volume form of $M_I^{(e)}$. We put $\tau(I)=dt$ for
$I=\emptyset$.

In accordance with the terminology of $p$-brane theory \cite{Stelle}
an $(n_a-1)$-form potential
\beq{2.26}
A^{(a,e,I)}=\Phi^{(a,e,I)}(u)\tau(I),\
\rank A^{(a,e,I)}\equiv n_a-1=d(I),\ a\in \Delta,
\eeq
where $\Phi^{(a,e,I)}(u)$ is a scalar function, describes an electrically
charged  $p$-brane ($p=n_a-2$)
with the worldvolume $M_I^{(e)}$.
Moreover,
the submanifold $\R_+\times S^{d_0}\times M_{\bar I}$
($\R_+\times S^{d_0}$ for $I=\{2,\ldots,n\}$)
is the so-called transverse space for this $p$-brane.
The $n_a$-form field
strength corresponding to $A^{(a,e,I)}$ was defined by
(\ref{2.3}) and may be written as
\beq{2.27}
F^{(a,e,I)}=d\Phi^{(a,e,I)}(u)\wedge\tau(I)
=\dot{\Phi}^{(a,e,I)}(u)du\wedge\tau(I).
\eeq
The overdot means a derivative with respect to the radial coordinate
$u$.

An  $n_b$-form field strength
\beq{2.28}
F^{(b,m,J)}=\e^{-2\lambda_b(\varphi)}
*\left(d\Phi^{(b,m,J)}(u)\wedge\tau(J)\right),\
J\in\Omega_0,\ b\in \Delta,
\eeq
describes a $p$-brane ($p=n_b-1=D-d(J) -2$) with a magnetic-type
charge.  The submanifold
\beq{2.29}
M_J^{(m)}=S^{d_0}\times M_{\bar J}
\eeq
is a worldvolume of this $p$-brane.
Clearly, $M_J^{(m)}=S^{d_0}$ for $J=\{2,\ldots,n\}$.
By $*$ we denoted the Hodge operator on the manifold $({\rm M},g)$, i.e.
\beq{2.30}
(*F)_{M_1\ldots M_{D-r}}=\frac{\sqrt{|g|}}{r!}
\eps_{N_1\ldots N_r M_1\ldots M_{D-r}}
F^{N_1\ldots N_r}.
\eeq

In this paper we consider the so-called composite $p$-branes
\cite{A2}, i.e., by definition we put
\beq{2.31}
F^a=\sum_{I\in\Omega_{a,e}}F^{(a,e,I)}+\sum_{J\in\Omega_{a,m}}F^{(a,m,J)},
\eeq
where $\Omega_{a,e}\subset\Omega_0$ is a subset
(which may be empty) of all $I\in\Omega_0$ such that
$d(I)+1=n_a\equiv\rank F^{(a,e,I)}$. Moreover,
$\Omega_{a,m}\subset\Omega_0$ is a subset (which may be
empty) of all $J\in\Omega_0$ such that
$D-d(J)-1=\dim M_J^{(m)}=n_a\equiv\rank F^{(a,m,J)}$.
Evidently,  $\Omega_{a,m}={\emptyset}$ for $n_a=D-1,D$.

We obtain the following non-zero
components of the Ricci tensor for the metric (\ref{2.14})
\bear{2.32}
R_0^0=-\e^{-2\gamma}\left( \sum_{k=0}^{n} d_{k}(\dot{x}^{k})^2+
\ddot{\gamma_0}- \dot{\gamma}\dot{\gamma_0}\right),\\
\label{2.33}
R_{n_k}^{m_k}=\left\{ \delta^k_0(d_0-1)\e^{-2x^k}
- \left[\ddot{x}^{k}+ \dot{x}^{k}(\dot{\gamma_0}-\dot{\gamma})
\right] \e^{-2\gamma} \right\} \delta_{n_k}^{m_k},
\ear
where we denoted
\beq{2.34}
\gamma_0=\sum_{k=0}^{n} d_{k}x^{k}.
\eeq
Indices $m_k$ and $n_k$ in (\ref{2.33}) for
$k=0,\ldots,n$ run over from ($D-\sum_{l=k}^{n}d_l$) to
($D-\sum_{l=k}^{n}d_l+d_k-1$).
We recall that $D=1+\sum_{k=0}^{n}d_k=\dim{\rm M}$.

Under the above assumptions related to the $F^a$-fields and the
metric (\ref{2.14})
the Maxwell-like equations (\ref{2.8}) and the Bianchi identities $dF^a=0$
have the following form, correspondingly
\bear{2.35}
&&\frac{d}{du}\left[ \e^{
\gamma_0 - \gamma - 2\sigma(I) + 2\lambda_a(\phi) }
\dot{\Phi}^{(a,e,I)}(u)\right]=0, \ I\in\Omega_{a,e},\ a\in\Delta\\
\label{2.36}
&&\frac{d}{du}\left[
\e^{ \gamma_0 - \gamma - 2\sigma(J) - 2\lambda_a(\phi) }
\dot{\Phi}^{(a,m,J)}(u)\right]=0,\ J\in\Omega_{a,m},\ a\in\Delta,
\ear
where
\beq{2.37}
\sigma(I)=d_1x^1+\sum_{i\in I}d_ix^i.
\eeq
For empty $I=\emptyset$ we put $\sigma(I)=d_1x^1$.

To denote  $F^a$-fields and their potentials,
it is useful the following collective index
\beq{2.38}
s=(a,v,I),\
I\in\Omega_{a,v},\ v=e,m,\ a\in\Delta.
\eeq
By $S$ we denote the set of
all elements $s$, i.e.
\beq{2.39} S=\bigsqcup
_{v=e,m}\left(\bigsqcup_{a\in\Delta}\{a\}\times\{v\}\times
\Omega_{a,v}\right).
\eeq

Integrating (\ref{2.35}) and (\ref{2.36}), we get
\beq{2.40}
\dot{\Phi}^s(u)=Q_s
\exp
\left[
\gamma - \gamma_0 + 2\sigma(I_s) - 2\chi_s\lambda_{a_s}(\phi)
\right],
\ s=(a_s,v_s,I_s)\in S,
\eeq
where
\bear{2.41}
\chi_s=+1,\ v_s=e,\\
\label{2.42}
\chi_s=-1,\ v_s=m.
\ear
$Q_s$  are arbitrary constants.

Let $S_*\subset S$ be a subset of all $s\in S$ such that
$Q_s\neq 0$.

To obtain the tensors $T[F^a]^M_N$ in a block-diagonal form,
we put the following restriction:
there are no elements $(a,v,I),(a,v,J)\in S_*$
such that
\beq{2.43}
I=(I\cap J)\sqcup
\{i\},\ J=(I\cap J)\sqcup \{j\},\ i\neq j,\ d_i=d_j=1,
\eeq
where $i,j=2,\ldots,n$.
Here the intersection $I\cap J$ may be empty.
The total energy-momentum tensor of  $F^a$-fields
has a block-diagonal form whenever the restriction (\ref{2.43}) is valid.
Using (\ref{2.40}), we present  its non-zero components in the form
\bear{2.44}
T[F^a]^0_0  = -\frac{1}{2} \e^{ -2 \gamma_0 }
\sum\limits_{s\in S_*} Q^2_s
\exp\left[ 2\sigma(I_s) - 2\chi_s\lambda_{a_s}(\phi) \right],\\
\label{2.45}
T[F^a]^{m_k}_{n_k}  = -\frac{1}{2} \e^{ -2 \gamma_0 }
\left(
\sum\limits_{s\in S_*}
(2\delta_{kI_s}-1)
Q^2_s
\exp
\left[ 2\sigma(I_s) - 2\chi_s\lambda_{a_s}(\phi) \right]
\right)
\delta^{m_k}_{n_k},
\ear
where
\beq{2.46}
\delta_{kI}=\sum_{i\in I}\delta_{ki}+\delta_{k1},
\ I\in\Omega_0,\ k=0,1,\ldots,n.
\eeq
We put $\delta_{kI}=\delta_{k1}$ for $I=\emptyset$. Evidently,
$\delta_{0I}=0$ and $\delta_{1I}=1$ for any $I\in\Omega_0$.

We assume that the dilatonic scalar fields $\varphi^{\alpha}$ depend
only on the radial coordinate $u$.
Under this assumption
the total energy-momentum tensor of the dilatonic scalar fields reads
\beq{2.47}
\left( T[\varphi]^M_N \right)=
\frac{1}{2}\e^{-2\gamma}\left(
\sum_{\alpha,\beta=1}^\omega h_{\alpha\beta}
\dot{\varphi}^{\alpha}\dot{\varphi}^{\beta}\right)
\diag (1,\underbrace{1,\ldots,1}_{d_0\ {\rm times}},-1,1,\ldots,1).
\eeq

The Einstein equations (\ref{2.6})
can be written as
$R^M_N=T^M_N-T\delta^M_N/(D-2)$.
Further we employ the equations
$R^{0}_{0}-R/2=T^{0}_{0}$ and
$R^{m_k}_{n_k}=T^{m_k}_{n_k}-
T\delta^{m_k}_{n_k}/(D-2)$.                                   \linebreak
Using
(\ref{2.32}),(\ref{2.33}),(\ref{2.44})(\ref{2.45}),(\ref{2.47}),
we obtain these equations in the form
\beq{2.48}
\frac{1}{2}\left(
\sum_{k,l=0}^{n}G_{kl}\dot{x}^k\dot{x}^l+
\sum_{\alpha,\beta=1}^\omega h_{\alpha\beta}
\dot{\varphi}^{\alpha}\dot{\varphi}^{\beta}\right)
+ V=0,
\eeq
\bear{2.49}
\ddot{x}^k + (\dot{\gamma_0}-\dot{\gamma})\dot{x}^k
&  = &
\e^{2\gamma}
\Biggl\{
\delta^k_0(d_0-1) \e^{-2x^k}\nonumber\\
& + &
\e^{-2\gamma_0}
\sum\limits_{s\in S_*} \chi_s
\left( \delta_{kI_s}-\frac{d(I_s)}{D-2}\right)Q^2_s
\exp
\left[ 2\sigma(I_s) - 2\chi_s\lambda_{a_s}(\phi) \right]
\Biggr\}
\ear
where we denoted
\beq{2.50}
G_{kl}=d_{k}\delta_{kl}-d_{k}d_{l},\ k,l=0,1,\ldots,n,
\end{equation}
\bear{2.51}
V=\frac{1}{2}\e^{2\gamma}\left\{
(d_0-1)d_0\e^{-2x^0}-
\e^{-2\gamma_0}
\sum\limits_{s\in S_*}Q^2_s
\exp
\left[ 2\sigma(I_s) - 2\chi_s\lambda_{a_s}(\phi) \right]
\right\}.
\ear
Under the above assumptions equations (\ref{2.7}) have the form
\bear{2.52}
\ddot{\varphi}^{\alpha} + (\dot{\gamma_0}-
\dot{\gamma})\dot{\varphi}^{\alpha}
=\e^{2\gamma-2\gamma_0}
\sum\limits_{s\in S_*}
\chi_s\lambda^{\alpha}_{a_s}Q^2_s
\exp
\left[ 2\sigma(I_s) - 2\chi_s\lambda_{a_s}(\phi) \right].
\ear

It is not difficult to verify that equations
(\ref{2.48}),(\ref{2.49}),(\ref{2.53}) may be presented as the
Euler-Lagrange equations obtained from the Lagrangian
\beq{2.53}
L=\e^{\gamma_0-\gamma}
\left[
\frac{1}{2}\left(\sum_{k,l=0}^{n}G_{kl}\dot{x}^{k}\dot{x}^{l}
+\sum_{\alpha,\beta=1}^\omega h_{\alpha\beta}
\dot{\varphi}^{\alpha}\dot{\varphi}^{\beta}
\right)
-V
\right]
\end{equation}
viewed as a function of
the generalized coordinates $\gamma,x^k,\varphi^\alpha$.
The equation \linebreak
$\partial L/\partial \gamma=d(\partial L/\partial
\dot{\gamma})/du$ leads to the zero-energy constraint (\ref{2.48}).
After the gauge fixing: $\gamma=F(x^k,\varphi^\alpha)$ the equations
(\ref{2.49}),(\ref{2.52}) may be considered as the Euler-Lagrange
equations obtained from the Lagrangian (\ref{2.53}) under the
constraint (\ref{2.48}).
Further, we use the so-called harmonic gauge
\beq{2.54}
\gamma\equiv
\gamma_0=\sum_{k=0}^nd_kx^k.
\eeq
It is easy to check that the radial coordinate $u$ is a harmonic
function in this gauge, i.e.  ${\btu}[g]u=0$.

Let us introduce an $(n+\omega+1)$-dimensional real vector
space $\R^{n+\omega+1}$. Denote by  $e_A,\ A=0,1,\ldots,n+\omega$,
the canonical basis in  $\R^{n+\omega+1}$ ($e_{1}=(1,0,\ldots,0$) etc.).
Define the following vectors:
\begin{enumerate}
\item
The vector whose coordinates are to be found
\beq{2.55}
x(u)=\sum_{A=0}^{n+\omega}x^A(u)e_A,\
\left(x^A(u)\right)=(x^0(u),
\ldots,x^n(u),\varphi^1(u),\ldots,\varphi^\omega(u)). \eeq
\item
The vector corresponding to the factor-space $M_0\equiv S^{d_0}$ with
a non-zero Ricci tensor. Hereafter, we call it by the vector induced by the
curvature of $M_0$
\beq{2.56}
V_0=\sum_{A=0}^{n+\omega}V^A_0e_A=-\frac{2}{d_0}e_0,\
\left(V_0^A\right)=-2\left(\frac{1}{d_0},0,\ldots,0\right).
\eeq
\item
The vector induced by a $p$-brane
\beq{2.57}
U_s=\sum_{A=0}^{n+\omega}U_s^Ae_A,\
\left(U_s^A\right)=2\left(
\underbrace{\delta_{kI}-d(I_s)/(D-2)}_{n+1},
\underbrace{-\chi_s\lambda_{a_s}^\alpha}_\omega
\right).
\eeq
\end{enumerate}
A set of the vectors $U_s$  characterizes
the space-time M, the configuration of $p$-branes and their couplings to
dilatonic scalar fields. Further these vectors are called characteristic.

Let $<.,.>$ be a symmetrical bilinear form on
$\R^{n+\omega+1}$ such that
\beq{2.58}
<e_A,e_B>=\bar G_{AB},
\eeq where we put
by definition
\bear{2.59}
(\bar G_{AB})=\left(\begin{array}{cc} G_{kl}&0\\
0&h_{\alpha\beta}
\end{array}\right).
\ear
The form is nondegenerate, the inverse matrix to
$(G_{AB})$ reads
\bear{2.60}
(\bar G^{AB})=\left(\begin{array}{cc}
\frac{\delta^{kl}}{d_{k}}+\frac{1}{2-D}  &0\\
0&h^{\alpha\beta}
\end{array}\right).
\ear
The form $<.,.>$ endows the space $\R^{n+\omega+1}$ with the metric,
whose signature is $(-, +,\ldots, +)$ \cite {I3}.
By the usual way we may introduce the covariant components of vectors.
For the vectors
$V_0, U_s$
the covariant components have the form
\bear{2.61}
V_{0,A}=
2(
\underbrace{d_k-\delta_{0k}}_{n+1},
\underbrace{0,\ldots,0}_\omega),\\
\label{2.62}
U_{s,A}=
2(\underbrace{d_k\delta_{kI}}_{n+1},
\underbrace{-\chi_s\lambda_{a_s,\alpha}}_\omega),
\ear
The values of the bilinear form $<.,.>$ for
$V_0, U_s$
look as follows
\bear{2.63}
&&<V_0,V_0>=-4\frac{d_0-1}{d_0},\\
\label{2.64}
&&<V_0,U_s>=0,\ \forall s\in S,\\
\label{2.65}
&&<U_s,U_{s'}>=
4\left[ d(I_s\cap I_{s'})-\frac{d(I_s)d(I_{s'})}{D-2}+
\chi_s\chi_{s'}
\sum_{\alpha,\beta=1}^\omega h_{\alpha\beta}\lambda_{a_s}^\alpha
\lambda_{a_{s'}}^\beta
\right],
\ear
where $s=(a_s,v_s,I_s), s'=(a_{s'},v_{s'},I_{s'})\in S$.

A vector $y\in R^{n+\omega+1}$ is called time-like, space-like
or isotropic, if $<y,y>$  has  negative,  positive  or  null values
respectively. Vectors $y$ and $z$ are called orthogonal if
$<y,z>=0$. It should be noted that the curvature induced vector
$V_0$ is always time-like, while the $p$-brane
induced vector $U_s$  admits any value of
$<U_s,U_s>$. We mention that $V_0$ and $U_s$ are always orthogonal.

Using the notation $<.,.>$ and the vectors  (\ref{2.55})-(\ref{2.57}),
we represent the Lagrangian (\ref{2.53})  and the zero-energy constraint
(\ref{2.48})  with respect to a harmonic time gauge in the form
\bear{2.66}
L=\frac{1}{2}<\dot x,\dot x>-V,\\
\label{2.67}
E=\frac{1}{2}<\dot x,\dot x>+V\equiv 0,
\ear
where the potential $V$ reads
\beq{2.68}
V=a^{(0)} \e^{<V_0,x>}+
\sum_{s\in S_*}a^{(s)} \e^{<U_s,x>}.
\eeq

The following notation is used
\bear{2.69}
&&a^{(0)}=\frac{d_0(d_0-1)}{2}, \
a^{(s)}=-\frac{1}{2}Q^2_s,\ s\in S_*.
\ear

>From the mathematical viewpoint the obtaining of exact solutions
in the $p$-brane model under consideration is
reduced to  integration  of  equations of motion
for a system
with $(n+\omega+1)$ degrees of freedom described by the Lagrangian of the
form
\beq{2.70} L= \frac{1}{2}<\dot{x},\dot{x}>- \sum_{\mu=1}^ra^{(\mu)}
\e^{<b_{\mu}, x>},
\eeq
where $x,b_{\mu}\in \R^{n+\omega+1}$.  It should be noted
that the kinetic term $<\dot{x},\dot{x}>$ is not a positively definite
quadratic form as there
 usually takes place in classical mechanics. Due to the
pseudo-Euclidean signature  $(-, +, ..., +)$ of the form $<.,.>$ such
systems may be called the pseudo-Euclidean Toda-like systems as the
potential like that given in (\ref{2.70}) defines the
algebraic generalizations of the Toda chain \cite{Toda}.
well-known in classical mechanics .
\section{\bf Integration of the $p$-brane model with linearly independent
characteristic vectors} \setcounter{equation}{0}

We recall that $S_*\subset S$ is the subset of all $s\in S$ such that
$Q_s\neq 0$. Define a bijection
$f: S_* \mapsto \{1,2,\ldots,r\}$, where we denote by $r$ the cardinal
number of $S_*$, i.e.
\beq{3.1}
r=\abs{S_*}.
\eeq
Denote the natural number $f(s)$ corresponding to $s\in S_*$ by the same
letter $s$.

The problem consists in integrating the equations of motion obtained from
the Lagrangian  (\ref{2.66}) under the zero-energy constraint
(\ref{2.67}). Suppose the characteristic vectors $U_s\in \R^{n+\omega+1}$,
induced by $p$-branes are linearly independent. Then $r\leq n+\omega$.
We introduce a basis
$\{f_A\}$ in $\R^{n+\omega+1}$ in the following manner
\bear{3.2}
f_0=\frac{V_0}{<V_0,V_0>},\
f_s=\frac{2U_s}{<U_s,U_s>},\ s=1,\ldots,r,\\
\label{3.3}
<f_A,f_p>=\delta_{Ap},\ A=0,\ldots,n+\omega, p=r+1,\ldots,n+\omega.
\ear
Notice that if $r\equiv\abs{S_*}=n+\omega$ then the basis $\{f_A\}$
does not contain the vectors $f_p$ with $p\geq r+1$. We also mention
that due to the relation (\ref{2.64}) we get $<f_0,f_s>=0$ for
$s=1,\ldots,r$. It is not difficult to prove that the vectors
$f_1,\ldots,f_r,f_{r+1},\ldots,f_{n+\omega}$ must be space-like.

Using the decomposition
\beq{3.4}
x(u)=q^0(u)f_0 + \sum_{s=1}^r [q^s(u)-\ln{C^s}]f_s+
\sum_{p=r+1}^{n+\omega}q^{p}(u)f_p,
\eeq
we present the Lagrangian (\ref{2.66}) and the
constraint (\ref{2.67}) in the form
\bear{3.5}
L=L_0+L_T+L_P,\\
\label {3.6}
E=E_0+E_T+E_P\equiv 0,
\ear
where
\bear{3.7}
&&L_0=\frac{(\dot{q}^0)^2}{2<V_0,V_0>}-\frac{d_0(d_0-1)}{2}\e^{q^0},\\
\label{3.8}
&&E_0=\frac{(\dot{q}^0)^2}{2<V_0,V_0>}+\frac{d_0(d_0-1)}{2}\e^{q^0},\\
\label{3.9}
&&L_T=\sum_{s,s'=1}^r\frac{C_{ss'}}{<U_s,U_s>}\dot q^{s}\dot
q^{s'}+ \sum_{s=1}^r\frac{2}{<U_s,U_s>}
\exp\left[\sum_{s'=1}^rC_{ss'}q^{s'}\right],\\
\label{3.10}
&&E_T=\sum_{s,s'=1}^r\frac{C_{ss'}}{<U_s,U_s>}\dot q^{s}\dot
q^{s'}- \sum_{s=1}^r \frac{2}{<U_s,U_s>}
\exp\left[\sum_{s'=1}^rC_{ss'}q^{s'}\right],\\
\label{3.11}
&&L_P=E_P=\frac{1}{2}\sum_{p=r+1}^{n+\omega}(\dot{q}^{p})^2.
\ear
We introduced, in (\ref{3.9}),(\ref{3.10}), the nondegenerate Cartan-type
matrix $(C_{ss'})$ by the following manner
\beq{3.12}
C_{ss'}=2\frac{ <U_s,U_{s'}> }{ <U_{s'},U_{s'}> },\
s,s'=1,\ldots,r.
\eeq
The constants $C^s$ in the decomposition (\ref{3.4}) are defined by
\beq{3.13}
C^s=\prod_{s'=1}^r \left[\frac{ <U_{s'},U_{s'}> }{4}Q^2_{s'} \right]
^{C^{ss'}},\ \ s=1,\ldots,r,
\eeq
where $(C^{ss'})$ is the inverse matrix to  $(C_{ss'})$.

The Euler-Lagrange equations for
$q^{r+1}(u),\ldots,q^{n+\omega}(u)$ read \linebreak
$\ddot{q}^{r+1}(u)=\ldots=\ddot{q}^{n+\omega}(u)=0$.
Integrating them, we get
\bear{3.14}
q^p(u)=a^pu+b^p,\ p=r+1,\ldots,n+\omega,\\
\label{3.15}
E_P=\frac{1}{2}\sum_{p=r+1}^{n+\omega}(a^p)^2\geq 0,
\ear
where the constants $a^p,b^p$ are arbitrary.

For $q^0(u)$ we get the Liouville equation. The result of its integration
reads
\beq{3.16}
\e^{-q^0(u)/2}=F_0(u-u_0),
\eeq
where $u_0$ is an arbitrary constant. The function $F_0$ is defined by
\beq{3.17}
F_0(u)=
\frac
{ \sin\left[ \sqrt{ \frac{2E_0}{d_0(d_0-1) } }(d_0-1)u\right]}
{ \sqrt{\frac{2E_0}{d_0(d_0-1)}}}.
\eeq
This representation implies
\bear{3.18}
F_0(u)&&=(d_0-1)u,\  E_0=0,\\
\label{3.19}
&&=
\frac
{ \sin\left[ \sqrt{ \frac{2E_0}{d_0(d_0-1) } }(d_0-1)u\right]}
{ \sqrt{\frac{2E_0}{d_0(d_0-1)}}},\ E_0>0\\
\label{3.20}
&&=
\frac
{ \sinh\left[ \sqrt{ \frac{2\abs{E_0}}{d_0(d_0-1) } }(d_0-1)u\right]}
{ \sqrt{\frac{2\abs{E_0}}{d_0(d_0-1)}}},\ E_0<0.
\ear

The equations of motion for $q^1(u),\ldots,q^r(u)$ look as follows
\beq{3.21}
\ddot{q^s}=\exp\left[\sum_{s'=1}^rC_{ss'}q^{s'}\right],\
s,=1,\ldots,r.
\eeq
Using the transformation
\beq{3.22}
F_s(u)=
\e^{-q^s(u)},
\eeq
we present the set of equations (\ref{3.21}) in the form
\beq{3.23}
\dot{F}_s^2-F_s\ddot{F}_s=F_s^2\prod_{s'=1}^r(F_{s'})^{-C_{ss'}}.
\eeq
The set of equations (\ref{3.21}) proved to be completely integrable if
$(C_{ss'})$ is the Cartan matrix of a simple complex Lie algebra. The
general solutions for some algebras as well as some particular solution of
the set (\ref{3.21}) for  quite a wide class of matrices $(C_{ss'})$
will be considered in the next sections. Here we suppose that the
functions are known and the corresponding integral of motion (\ref{3.10})
is calculated.

Combining
(\ref{3.2}),(\ref{3.14}),(\ref{3.16})(\ref{3.22}), we present the
decomposition (\ref{3.4}) in the following form
\beq{3.24}
x(u)=-\ln[F_0(u-u_0)]\frac{2V_0}{<V_0,V_0>}
-\sum_{s=1}^r\ln[C^s F_s(u)]\frac{2U_s}{<U_s,U_s>}+
uQ+P,
\eeq
where  vectors $Q,P\in \R^{n+\omega+1}$ are defined by
\beq{3.25}
Q=\sum_{p=r+1}^{n+\omega}a^pf_p\equiv \sum_{A=0}^{n+\omega}Q^Ae_A,\
P=\sum_{p=r+1}^{n+\omega}b^pf_p\equiv \sum_{A=0}^{n+\omega}P^Ae_A,
\eeq
Due to the assumptions (\ref{3.3}) their coordinates
$Q^A,P^A$ w.r.t. the canonical basis $\{e_A\}$satisfy the constraints
\bear{3.26}
<Q,V_0>=2\sum_{k=0}^nQ^k(d_k-\delta_{k0})=0,\
<P,V_0>=2\sum_{k=0}^nP^k(d_k-\delta_{k0})=0.\\
\label{3.27}
<Q,U_s>=\sum_{A=0}^{n+\omega}Q^AU_{s,A}=0,\
<P,U_s>=\sum_{A=0}^{n+\omega}P^AU_{s,A}=0,\ s=1,\ldots,r.
\ear

Finally, the exact solution can be summarized as follows.
\begin{enumerate}
\item
The metric (\ref{2.14}) in the harmonic time gauge (\ref{2.54})
reads
\bear{3.28}
& ds^2 &
=
\prod\limits_{s=1}^r
\left[C^sF_s(u)\right]
^{\frac{8d(I_s)}{(D-2)<U_s,U_s>}}
\biggl\{
[F_0(u-u_0)]^{\frac{2}{1-d_0}}
\e^{2Q^0u+2P^0}
\left[F_0^{-2}(u-u_0)du^2+d\Omega_{d_0}^2\right]
\nn\\
&-&
\prod\limits_{s=1}^r
\left[C^sF_s(u)\right]
^{\frac{-8}{<U_s,U_s>}}
\e^{2Q^1u+2P^1}
dt^2
+\sum\limits_{i=2}^n
\prod\limits_{s=1}^r
\left[C^sF_s(u)\right]
^{\frac{-8\delta_{iI_s}}{<U_s,U_s>}}
\e^{2Q^iu+2P^i}
ds_i^2
\biggr\}.
\ear
\item
The dilatonic scalar fields are the following
\beq{3.29}
\varphi^\alpha(u)=\sum_{s=1}^r
\frac{4\chi_s\lambda^\alpha_{a_s}}{<U_s,U_s>}
\ln[C^sF_s(u)]
+uQ^{n+\alpha}+P^{n+\alpha},\ \alpha=1,\ldots,\omega.
\eeq
\item
For scalar functions $\dot{\Phi}^s(u)$
we get
\beq{3.30}
\dot{\Phi}^s(u)=Q_s\e^{<U_s,x>}=
Q_s\prod_{s'=1}^m \left[C^{s'} F_{s'}(u)\right]^{-C_{ss'}}
,\ s=1,\ldots,r.
\eeq
\end{enumerate}
The corresponding $F^a$-field forms look as follows
\beq{3.31}
F^{(a_s,e,I_s)}=\dot{\Phi}^{(a_s,e,I_s)}du\wedge\tau(I_s)
\eeq
for the electrically charged $p$-brane and
\beq{3.32}
F^{(a_s,m,I_s)}=Q_s\tau_0\wedge\tau_{i_1}\wedge\ldots
\wedge\tau_{i_c},\ \{i_1,\ldots,i_c\}=\bar{I_s},\ c=n_{a_s}-d_0
\eeq
for the $p$-brane with magnetic-type charge. We put
$F^{(a_s,m,I_s)}=Q_s\tau_0$ if $\bar{I_s}=\emptyset$.
We stress that if $r\equiv |S_*|=n+\omega$, then one must put
$Q^A=P^A=0,\ A=0,\ldots,n+\omega$ in this solution.

\section{\bf General solutions for models
associated with Lie algebras} \setcounter{equation}{0}
Now we list general solutions to the set of equations (\ref{3.23})
for some special matrices $(C_{ss'})$.
\begin{enumerate}
\item
$(C_{ss'})=\diag(2,\ldots,2)$ is the Cartan matrix of the semi-simple
Lie  algebra $A_1\oplus\ldots\oplus A_1$ of  rank $r$. In this case the
set of the characteristic vectors $U_s$ is orthogonal.

\bear{4.1}
F_s(u)\equiv \e^{-q^s(u)}=\frac{\sin[w_s(u-u_{01})]}{w_s},\
s=1,\ldots,r,\\
\label{4.2}
E_T=\sum_{s=1}^r\frac{2w_s^2}{<U_s,U_s>},
\ear
where $w_s$ are arbitrary constants, which may be real (including zero)
or imaginary.
\item
$(C_{ss'})=(2\delta_{ss'}-\delta_{s,s'+1}-\delta_{s,s'-1})$ is the Cartan
matrix of the simple Lie algebra $A_r\equiv sl(r+1,C)$.  In this case
all characteristic vectors $U_s$ are space-like with coinciding lengths,
i.e.
\beq{4.3}
<U_s,U_s>\equiv U^2,\ s=1,\ldots,r.
\eeq
By the transformation
\beq{4.4}
q^s\mapsto q^s-\frac{\pi i}{2}m_s,
\eeq
where
\beq{4.5}
m_s=2\sum_{s'=1}^r C^{ss'}=s(r+1-s),
\eeq
we put the set of equations (\ref{3.21}) into the form
\beq{4.6}
\ddot{q^s}=-\exp\left[\sum_{s'=1}^rC_{ss'}q^{s'}\right],\
s,=1,\ldots,r.
\eeq
These are precisely the $A_r$ Toda equations \cite{Toda}. Using the
general solutions to these equations presented  by
Anderson \cite{Anderson}, we obtain the following result
\bear{4.7}
&&F_s(u)= i^{s(r+1-s)}
\sum_{\mu_1<\ldots<\mu_s}^{r+1} v_{\mu_1}\cdots
v_{\mu_s} \Delta^2(\mu_1,\ldots,\mu_s)\e^{(w_{\mu_1}+\ldots
+w_{\mu_s})u},\\
\label{4.8}
&&E_T=\frac{1}{2}\sum_{\mu=1}^{r+1} w^2_\mu.
\ear
 where $\Delta^2(\mu_1,\ldots,\mu_s)$ denotes the square of
the Vandermonde determinant
\beq{4.9}
\Delta^2(\mu_1,\ldots,\mu_s)=\prod_{\mu_i<\mu_j}
\left(w_{\mu_i}-w_{\mu_j}\right)^2,\
\Delta^2(\mu_1)\equiv 1.
\eeq
The constants $v_\mu$ and $w_m,\ \mu=1,\ldots,r+1$, have to satisfy the
following constraints:
\beq{4.10}
\prod_{\mu=1}^{r+1}
v_\mu=\Delta^{-2}(1,\ldots,r+1), \ \ \ \sum_{\mu=1}^{r+1}w_\mu=0\; .
\eeq
The constants $v_\mu,w_\mu$ are in general complex. There are additional
constraints on them if one requires the functions $F_s(u)$ and the
integral of motion (\ref{4.8}) to be real. In (\ref{4.7}) we used
$m_s=s(r+1-s)$ for $A_r$.

\item
$(C_{ss'})$ is the following matrix
\bear{4.11}
C_{ss'}=\left\{
\begin{array}{lcl}
2\delta_{ss'}-\delta_{s,s'+1}-\delta_{s,s'-1} &{\rm for}&
s=1,\ldots,r,\ s'=1,\ldots,r-1,\\
\delta_{ss'}-\delta_{s,s'-1} &{\rm for}&
s=1,\ldots,r,\ s'=r.\\
\end{array}
\right.
\ear
The Cartan matrix of the simple Lie algebra $B_r\equiv so(2r+1)$ may be
obtained from that given in (\ref{4.11}) by multiplying the last column
of $(C_{sr})$ by 2. In this case the general solution to the set of
equations (\ref{3.21}) may be obtained from the previous formulae
(\ref{4.7}), (\ref{4.8}) as in \cite{Leznov}. Notice
that the equations (\ref{3.21})  are symmetric under the following
permutations $q^s\leftrightarrow q^{r+1-s}$ for $s=1,\ldots,r$ if
$(C_{ss'})$ if the Cartan matrix of $A_r$. This implies that there are
solutions (\ref{4.7}) with $q^s\equiv q^{r+1-s}$ for $s=1,\ldots,r$.
Moreover, this identification for $r=4,6,8,\ldots$ leads to the $(r/2)$
equations of the form (\ref{3.21}) with the matrix (\ref{4.11}).
Consequently the general solution of the equations (\ref{3.21}) with the
matrix (\ref{4.11}) for $r=r_0$ may be obtained form (\ref{4.7}) for
$r=2r_0$ by putting additional constraints on the constants $v_\mu,w_\mu$
providing with the identities $F_s(u)\equiv F_{2r_0+1-s}(u),\
s=1,\ldots,2r_0$.
\item $(C_{ss'})$ is the Cartan matrix of the simple Lie
algebra $C_r\equiv sp(r,C)$, i.e.
\bear{4.12}
C_{ss'}=\left\{
\begin{array}{lcl}
2\delta_{ss'}-\delta_{s-1,s'}-\delta_{s-1,s'} &{\rm for}&
s=1,\ldots,r-1,\ s'=1,\ldots,r,\\
\delta_{ss'}-\delta_{s,s'-1} &{\rm for}&
s=r,\ s'=1,\ldots,r.\\
\end{array}
\right.
\ear
In this case the general solution to the set of equations (\ref{3.21})
with $r=r_0$ may be obtained from (\ref{4.7}) with $r=2r_0-1$ by putting
additional constraints on the constants $v_\mu,w_\mu$
providing with
identities $F_s(u)\equiv F_{2r_0-s}(u),\ s=1,\ldots,2r_0-1$. It stems
>from the following property of the set (\ref{3.21}):
the identification $q^s\equiv q^{2r_0-s}$ for $s=1,\ldots,2r_0-1$
reduces the set  (\ref{3.21}) with the Cartan matrix of $A_{2r_0-1}$ to the
set (\ref{3.21}) with the Cartan matrix of $C_{r_0}$ ($r_0\geq 2$).

\end{enumerate}

\section{\bf The particular solution describing
black holes} \setcounter{equation}{0}

Suppose the nondegenerate Cartan-type matrix $(C_{ss'})$ satisfies the
conditions
\beq{5.1}
m_s=2\sum_{s'=1}^r C^{ss'}>0,\ s=1,\ldots,r.
\eeq
The conditions are valid for extremely large class of the $p$-brane models.
For instance, the parameters $m_s$ are natural numbers
if $(C_{ss'})$ is the Cartan
matrix of a semi-simple Lie algebra $G$ \cite{I9}. For $G=A_r=sl(r+1,C)$
the parameters $m_s$ are given by (\ref{4.5}).

Under the conditions (\ref{5.1}) the set of equations (\ref{3.23})
admits the following particular solution
\beq{5.2}
F_s(u)=a_s\left(\frac{\sinh[\bar\mu(u-u_{01})]}{\bar\mu}\right)^{m_s},\
s=1,\ldots,r,
\eeq
where the constants $a_s$ are defined by
\beq{5.3}
a_s=\prod_{s'=1}^r(m_{s'})^{-C^{ss'}}
\eeq
and the constants $\bar\mu,\ u_{01}$ are arbitrary. The corresponding to
(\ref{5.2}) integral of motion (\ref{3.10}) has the form
\beq{5.4}
E_T=\sum_{s,s'=1}^r\frac{C_{ss'}}{<U_s,U_s>}\frac{\dot F_s}{F_s}
\frac{\dot F_{s'}}{F_{s'}}-
\sum_{s=1}^r\frac{2}{<U_s,U_s>}\prod_{s'=1}^r(F_{s'})^{-C_{ss'}}=
2\bar\mu^2\sum_{s=1}^r\frac{m_s}{<U_s,U_s>}.
\eeq
For $\bar\mu=0$ the formulas (\ref{5.2}),(\ref{5.4}) read
\bear{5.5}
F_s(u)&=&a_s(u-u_{01})^{m_s},\ s=1,\ldots,r,\\
E_T&=&0.
\ear
It is evident that (\ref{5.5}) represents the polynomials in the radial
coordinate $u$ if the parameters $m_s$ are natural numbers.
As we have already
mentioned, $m_s$ are natural numbers
if, for instance, $(C_{ss'})$ is the Cartan
matrix of a semi-simple Lie algebra. The formula (\ref{5.5}) does not
exhaust all possible polynomial solutions to the set (\ref{3.23}).
As far as we know, an explicit general form of the polynomial solution,
which appears for the set of equations (\ref{3.23}) with arbitrary
natural numbers $m_s$ and vanishing integral of motion (\ref{5.4}),
is not found. There are only  few examples in the literature. For instance,
in \cite{Lu3} all possible polynomial solutions were obtained for the
matrix $(C_{ss'})$ supposed to be the Cartan matrix of the Lie algebras
$A_r\equiv sl(r+1,C)$ with $r=1,2,3$. It is easy to check that an
arbitrary polynomial solution to (\ref{3.23}) under the condition $E_T=0$
may be obtained from (\ref{5.5}) by adding some polynomial of lower degree
to the leading term $a_su^{m_s}$.

Now we use the particular solution (\ref{5.2}) with $\bar\mu > 0$ and its
special form (\ref{5.5}) corresponding to $\bar\mu=0$ for constructing
non-extremal and extremal black holes, respectively.
Consider the general form of exact solution (\ref{3.28})-(\ref{3.31})
under the following additional assumptions
\begin{enumerate}
\item
We put
\beq{5.7}
Q^A=\bar\mu\left(
\frac{2V_0^A}{<V_0,V_0>}+\sum_{s=1}^r\frac{2m_sU_s^A}{<U_s,U_s>}
-\delta_1^A\right),\ A=0,\ldots,n+\omega.
\eeq
One may verify that
the conditions $<Q,V_0>=0,\ <Q,U_s>=0,\ s=1,\ldots,r$ are valid.
Using the zero-energy constraint (\ref{3.6}) and (\ref{5.4}), we find
the constant $E_0$
\beq{5.8}
E_0=-E_T-\frac{1}{2}<Q,Q>=-\frac{1}{2}d_0\bar d_0\mu^2,
\eeq
where we denoted
\beq{5.9}
\mu=\bar\mu/\bar d_0,\ \bar d_0=d_0-1.
\eeq
\item
We take the parameters $P^A$ in the form
\beq{5.10}
\left(P^A\right)=
\sum_{s=1}^r\ln[C^sF_s(u_0)]\frac{2\left(U_s^A\right)}{<U_s,U_s>}-
u_0\left(Q^A\right)+
\ln\eps_0(1,0,\underbrace{-R^2,\ldots,R^n}_{n-1},
\underbrace{0,\ldots,0}_\omega),
\eeq
where $\eps_0$ is an arbitrary positive constant. The conditions
$<P,V_0>=0$ lead to the following constraint on parameters
$R^2,\ldots,R^n$
\beq{5.11}
\sum_{i=2}^nR^id_i=\bar d_0.
\eeq
Combining the conditions $<P,U_s>=0, s=1,\ldots,r$ and (\ref{3.13}),
we get
\beq{5.12}
Q_s^2=\frac{<U_s,U_s>}{4}\eps_0^{-2\sum_{i=2}^nd_i\delta_{iI_s}R^i}
\prod_{s'=1}^r[F_{s'}(u_0)]^{-C_{ss'}},\ s=1,\ldots,r.
\eeq
\item
Now we consider the solution for $u\in(u_0,+\infty)$ and introduce the
following new radial coordinate
\beq{5.13}
R=\frac{R_0}{1-\exp[-2\bar\mu(u-u_0)]}=
\eps_0\left(\frac{2\mu}{1-\exp[-2\bar\mu(u-u_0)]}\right)^{1/\bar d_0},\
R>R_0.
\eeq
The constant $R_0$ is defined by
\beq{5.14}
R_0=\eps_0(2\mu)^{1/\bar d_0}.
\eeq
Here we take the constant $u_{01}$ in (\ref{5.2}) such that
$(u_0-u_{01})>0$. Moreover we introduce the constant
\beq{5.15}
R_*=R\mid_{u=2u_0-u_{01}}=
\eps_0\left(\frac{2\mu}
{1-\exp[-2\bar\mu(u_0-u_{01})]}\right)^{1/\bar d_0}>R_0.
\eeq
\end{enumerate}

Finally, we obtain the metric
\bear{5.16}
ds^2&=&
\left[1+\left(\frac{R_*}{R}\right)^{\bar d_0}-
\left(\frac{R_0}{R}\right)^{\bar d_0}\right]
^{\sum_{s=1}^r\frac{8m_sd(I_s)}{(D-2)<U_s,U_s>}}
\Biggl\{\Biggr.
\frac{dR^2}{1-(R_0/R)^{\bar d_0}}+
R^2d\Omega^2_{d_0}\nonumber\\
&-&
\left[1+\left(\frac{R_*}{R}\right)^{\bar d_0}-
\left(\frac{R_0}{R}\right)^{\bar d_0}\right]
^{-\sum_{s=1}^r\frac{8m_s}{<U_s,U_s>}}
\left(1-\left(\frac{R_0}{R}\right)^{\bar d_0}\right)dt^2\nonumber\\
&+&
\sum_{i=2}^n\eps_0^{-2R^i}
\left[1+\left(\frac{R_*}{R}\right)^{\bar d_0}-
\left(\frac{R_0}{R}\right)^{\bar d_0}\right]
^{-\sum_{s=1}^r\frac{8m_s\delta_{iI_s}}{<U_s,U_s>}}ds_i^2
\Biggl.\Biggr\},
\ear

the dilatonic scalar fields
\beq{5.17}
\varphi^\alpha=
\sum_{s=1}^r\frac{4m_s\chi_s\lambda^\alpha_{a_s}}{<U_s,U_s>}
\ln
\left[1+\left(\frac{R_*}{R}\right)^{\bar d_0}-
\left(\frac{R_0}{R}\right)^{\bar d_0}\right],\ \alpha=1,\ldots,\omega,
\eeq

and the potential derivatives
\bear{5.18}
\frac{d\Phi^s}{dR}=&-&{\rm sgn}(Q_s)
2\bar d_0
\sqrt{\frac{m_s}{<U_s,U_s>}
\left(1-\left(\frac{R_0}{R_*}\right)^{\bar d_0}\right)}\nonumber\\
&\times&\eps_0^{-\sum_{i=2}^nd_i\delta_{iI_s}R^i}
\left(\frac{R_*}{R}\right)^{\bar d_0}
\left[1+\left(\frac{R_*}{R}\right)^{\bar d_0}-
\left(\frac{R_0}{R}\right)^{\bar d_0}\right]^{-2}.
\ear

The corresponding $F^a$-field forms may be obtained by
(\ref{3.31}),(\ref{3.32}), where
\beq{5.19}
\abs{Q_s}=
2\bar d_0
\sqrt{\frac{m_s}{<U_s,U_s>}
\left(1-\left(\frac{R_0}{R_*}\right)^{\bar d_0}\right)}
\eps_0^{\sum_{i=2}^nd_i\delta_{iI_s}R^i}
\left(\frac{R_*}{\eps_0}\right)^{\bar d_0}.
\eeq

Then constants $\bar\mu,\eps_0,(u_0-u_{01})$ are independent.
The constants $\mu,\bar d_0,R_0,R_*$ are defined by
(\ref{5.9}),(\ref{5.14}),\linebreak
(\ref{5.15}). The parameters $R^2,\ldots,R^n$
obey the relation (\ref{5.11}).

Now we analyze the particular solution (\ref{5.16})-(\ref{5.18}).
The metric (\ref{5.16}) is asymptotically flat, i.e.
\beq{5.20}
\lim_{R\to+\infty}ds^2=dR^2+R^2d\Omega^2_{d_0}-dt^2+\sum_{i=2}^n
\eps_0^{-2R^i}ds_i^2.
\eeq
According to (\ref{5.11}) all parameters $R^2,\ldots,R^n$ may be positive.
Then, the constant scale factors $\eps_0^{-2R^i}$ of internal spaces
$M_2,\ldots,M_n$ are arbitrary small if $\eps_0$ is large enough.

If $\bar\mu>0$  the particular solution describes a non-extremal black hole
with the horizon at $R=R_0$. The active gravitational mass $M_g$ and the
Hawking temperature $T_H$ of this black hole read
\bear{5.21}
&&2G_NM_g=R_*^{\bar d_0}
\left[
\left(1-\left(\frac{R_0}{R_*}\right)^{\bar d_0}\right)
\sum_{s=1}^r\frac{4m_sU_s^1}{<U_s,U_s>}+
\left(\frac{R_0}{R_*}\right)^{\bar d_0}
\right],\\
&&T_H=\frac{\bar d_0}{4\pi k_BR_*}\left(\frac{R_0}{R_*}\right)
^{\sum_{s=1}^r\frac{4m_s}{<U_s,U_s>}-1},
\ear
where $G_N$ and $k_b$ are Newton's gravitational constant and Boltzmann's
constant, respectively.

The solution (\ref{5.16})-(\ref{5.18}) may be considered in the so-called
extreme case, when $\bar\mu=0$ ($R_0=0$). It follows from general
statements proved in \cite{BIM} that the point $R=0$ is a curvature
singularity of the metric (\ref{5.16}) with $R_0=0$ if $T_H\to+\infty$
as $R_0\to +0$. Then, the particular solution admits an extremal black hole
only if
\beq{5.23}
\sum_{s=1}^r\frac{4m_s}{<U_s,U_s>}\geq 1.
\eeq


\begin{thebibliography}{99}


\bibitem{Stelle}
{\it K.S. Stelle}. "Lectures on Supergravity p-branes ",
hep-th/9701088.

\bibitem{Lu1}
{\it M.J. Duff, H. L\"u, and C.N. Pope}. "The Black Branes of
M-theory", hep-th/9604052.


\bibitem{Lu2}
{\it H. L\"u, C.N. Pope,  K.W. Xu}. "Liouville and Toda Solitons in
M-theory", hep-th/9604058.

\bibitem{Lu3}
{\it H. L\"u, and C.N. Pope}. "$SL(N+1,R)$ Toda Solitons in
Supergravities", hep-th/9607027.

\bibitem{Lu4}
{\it H. L\"u, and C.N. Pope}. "Black $p$-branes and Their Vertical
Dimensional Reduction", hep-th/9609126.



\bibitem{CT}
{\it M. Cvetic,  A.A. Tseytlin}.
Nucl. Phys. B. 1996. V. 478. P. 181.


\bibitem{I1}
{\it V.D. Ivashchuk,  V.N. Melnikov}.
"Intersecting p-brane Solutions in Multidimensional
Gravity and M-theory", hep-th/9612089;
Gravitation \& Cosmology. 1996. V. 2. P. 297.

\bibitem{A1}
{\it I.Ya. Aref'eva, O.A. Rytchkov}.
"Incidence Matrix Description of Intersecting p-brane
Solutions", hep-th/9612236.


\bibitem{A2}
{\it I.Ya. Aref'eva, M.G. Ivanov, O.A. Rytchkov}.
"Properties of Intersecting p-branes in Various Dimensions",
hep-th/9702077.

\bibitem{A3}
{\it I.Ya. Aref'eva, M.G. Ivanov,  I.V. Volovich}.
"Non-extremal Intersecting p-branes  in Various Dimensions",
hep-th/9702079.

\bibitem{Oh1}
{\it N. Ohta and T. Shimuzu}. " Non-extreme Black Holes from Intersecting
M-branes", hep-th/9701095.

\bibitem{Oh2}
{\it N. Ohta}. "Intersection Rules for Non-extreme p-branes",
hep-th/9702164.

\bibitem{I2}
{\it V.D. Ivashchuk, V.N. Melnikov}.
Phys. Lett. B. 1997. V. 403. P. 23.

\bibitem{Maldacena}
{\it J.M. Maldacena}. "Black Holes and D-branes", hep-th/9705078.

\bibitem{I3}
{\it V.D. Ivashchuk, V.N. Melnikov}. "Multidimensional Classical
and Quantum Cosmology with Intersecting $p$-branes",
hep-th/9708157;  J. Math. Phys. 1998. V. 39. P. 2866.


\bibitem{I4}
{\it V.D. Ivashchuk, V.N. Melnikov}.
"Sigma-model for the Generalized  Composite p-branes",
hep-th/9705036; Class. Quantum Grav. 1997. V. 14. P. 3001.

\bibitem{BIM}
{\it K.A. Bronnikov, V.D. Ivashchuk,  V.N. Melnikov}.
"The Reissner-Nordstr\"om Problem for
Intersecting Electric and Magnetic $p$-branes", gr-qc/9710054;
Gravitation \& Cosmology. 1997. V. 3. P. 203.


\bibitem{GrI}
M.A. Grebeniuk and V.D. Ivashchuk,
"Sigma-model Solutions and Intersecting p-branes Related to Lie Algebras",
 hep-th/9805113; Phys. Lett. B 1998. V. 442.P.125.


\bibitem{BKR}
{\it K.A. Bronnikov, U. Kasper,  M. Rainer}. "Intersecting Electric and
Magnetic $p$-branes: Spherically Symmetric Solutions", gr-qc/9708058;
Gen. Relativity and Gravitation. 1999. V.31. P.1681.

\bibitem{Br}
{\it K.A. Bronnikov}, "Block-orthogonal Brane Systems, Black Holes and
Wormholes", hep-th/9710207; Gravitation \& Cosmology. 1998.
V. 4. P. 49.

\bibitem{I5}
{\it V.D. Ivashchuk, V.N. Melnikov}. "Madjumdar-Papaptrou Type Solutions in
Sigma Model and Intersecting $p$-branes", hep-th/9802121;
Class. Quantum Grav. 1999. V. 16. P. 849.


\bibitem{I6}
{\it V.D. Ivashchuk,S.-W. Kim,  V.N. Melnikov}. "Hyperbolic Kac-Moody
Algebra from Intersecting $p$-branes", hep-th/9803006;
J. Math Phys. 1999.V.40. P. 4072.

\bibitem{I7}
{\it V.D. Ivashchuk and V.N. Melnikov}.
"Multidimensional Cosmological and Spherically Symmetrical Solutions
with Intersecting p-branes", gr-qc/9901001;
In: Lecture Notes in Physics, V. 537. Mathematical and Quantum Aspects
Of Relativity and Cosmology. Eds. S. Cotsakis and G. Gibbons.
Proc. 2nd Samos Meeting
(September 1998, Pithagoreon), Springer, Berlin, 2000. P. 214.

\bibitem{I8}
{\it S. Cotsakis, V.D. Ivashchuk and V.N. Melnikov}.
"p-brane Black Holes and Post-Newtonian Approximation", hep-th/9902148;
Gravitation \& Cosmology. 1999.  V. 5. P. 52.


\bibitem{I9}
{\it V.D. Ivashchuk and V.N. Melnikov}.
$p$-brane Black Hole Solution for General Intersection Rules",
hep-th/9910041; Gravitation \& Cosmology. 1999. V.5. P. 313.


\bibitem{Cremmer}
{\it E. Cremmer, B. Julia,  J. Scherk}.
Phys. Lett. B. 1978. V. 76. P. 409.

\bibitem{G}
{\it V.R. Gavrilov, V.D. Ivashchuk, V.N. Melnikov}
J. Math. Phys. 1995. V. 36. P. 5829.


\bibitem{Toda}
{\it M. Toda}. Theory of Nonlinear Lattices. Springer-Verlag,
Berlin. 1981

\bibitem{Anderson}
{\it A. Anderson}.   J. Math. Phys. 1996. V. 37. P. 1349.

\bibitem{Leznov}
{\it A.N. Leznov and M.V. Saveliev}. Group-theoretical Methods for
Integration of Nonlinear Dynamical Systems. Birkhauser, Basel, 1992.

\end{thebibliography}
\end{document}